\documentclass[fleqn,10pt]{wlscirep}
\usepackage[utf8]{inputenc}
\usepackage[T1]{fontenc}
\usepackage{subcaption}

\title{Impact of geography on the importance of parameters in infectious disease models}

\author[1,*]{Arindam Saha}
\author[1]{Maziar Ghorbani}
\author[1]{Diana Suleimenova}
\author[1]{Anastasia Anagnostou}
\author[1]{Derek Groen}
\affil[1]{Department of Computer Science, Brunel University London, The United Kingdom}

\affil[*]{arindamsaha1507@gmail.com}

\begin{abstract}
Agent-based models are widely used to predict infectious disease spread. For these predictions, one needs to understand how each input parameter affects the result. Here, some parameters may affect the sensitivities of others, requiring the analysis of higher order coefficients through e.g. Sobol sensitivity analysis. The geographical structures of real-world regions are distinct in that they are difficult to reduce to single parameter values, making a unified sensitivity analysis intractable. Yet analyzing the importance of geographical structure on the sensitivity of other input parameters is important because a strong effect would justify the use of models with real-world geographical representations, as opposed to stylized ones.

Here we perform a grouped Sobol's sensitivity analysis on COVID-19 spread simulations across a set of three diverse real-world geographical representations. We study the differences in both results and the sensitivity of non-geographical parameters across these geographies. By comparing Sobol indices of parameters across geographies, we find evidence that infection rate could have more sensitivity in regions where the population is segregated, while parameters like recovery period of mild cases are more sensitive in regions with mixed populations. We also show how geographical structure affects parameter sensitivity changes over time.

\end{abstract}

\begin{document}

\flushbottom
\maketitle
%
%
\thispagestyle{empty}


\section*{Introduction}

Modeling and prediction of infectious diseases play an important role in mitigating their spread and impact. Many models have been developed in the past to study the spread of infectious diseases such as the 2014 Ebola outbreak in Africa~\cite{Chretien2015,Rivers2014}, the 2009 H1N1 pandemic~\cite{Nishiura2010,Chowell2011,Chowell2012} and the 2015 Zika outbreak in South America\cite{OReilly2018}. Needless to say, the COVID-19 pandemic which has severely impacted the entire world has also been the subject of a significantly high number of modeling studies~\cite{Rahimi2021,Wynantsm2020,Adiga2020} given the severity of the pandemic and increased computational capabilities available now.

Infectious disease models can be broadly classified into two categories: differential equation models and agent-based models~\cite{Parunak1998,Brauer2008}. Differential equation models are based on the assumption that the population is well mixed and the disease spreads homogeneously across the population. These models are computationally efficient and can be easily used to study the spread of the disease in a large population in real-time. However, these models are not able to capture the effect of the geographical distribution of the population on the spread of the disease, which is necessary in realistic scenarios. Agent-based models, on the other hand, are based on the assumption that the population is not well-mixed and the disease spreads heterogeneously across the population. Therefore, they simulate the movement and mutual interactions of individual agents in the population. These models are computationally expensive and, without the use of high-performance computers (HPC's), can only be used to study the spread of the disease in a small population in real-time. During the COVID-19 crisis, many differential equation models~\cite{Leontitsis2021,Abou-Ismail2020,Kong2022} and agent-based models~\cite{Shamil2021, Cuevas2020, Kerr2021, Bullock2021, Ferguson2006, Ozik2021} have been developed and used to study the evolution of the pandemic.

The accuracy of any model depends on the assumptions made and the parameters used in it. Since agent-based models provide a more detailed representation of the population, they usually depend on a larger set of parameters compared to differential equation models. Therefore, it is important to understand how input assumptions and parameters affect the predictions of the model. This can be done by performing sensitivity analysis. 

In recent years, sensitivity analysis has become a popular tool for analysis of models due to its importance in a variety of fields. In any model, it is essential to quantify the uncertainties in the predictions it makes. In general, the model inputs are subject to sources of uncertainty, including errors of measurement, absence of information and poor or partial understanding of the driving forces and mechanisms. While such uncertainties in the inputs of a model can mostly be quantified, mathematically computing the uncertainties in the output is mostly impossible due to the complexities of the model. In such circumstances, sensitivity analysis becomes a useful tool to estimate the uncertainties and confidence intervals of the model outputs. Such applications of sensitivity analysis have been used in multiple fields of study\cite{Saltelli2020,Gelesz2020,Turanyi1997,Heiselberg2009}. Sensitivity analysis can also be used to assist model development itself as demonstrated by sensitivity-driven simulation development approach~\cite{Suleimenova2021}. Sensitivity analysis can be used in combination with validation techniques to identify the most important parameters in the model and iteratively steer the parameters to improve the model. This technique has been recently applied to improve the predictions of a human migration model~\cite{Suleimenova2017}.

Sensitivity analysis is a well-established field of study, and many methods have been developed to perform sensitivity analysis of models~\cite{Frey2002,Iooss2015,Hamby1995,Borgonovo2016}. These methods study how the uncertainty in the output of a model can be apportioned to different sources of uncertainty in their inputs. Essentially, sensitivity analysis provides a quantitative tool that can be used to sequentially order the parameters according to their importance in the model output. This can help reduce the computational cost of the model by reducing the number of parameters that need to be estimated. These methods can be broadly classified into two categories: local sensitivity analysis and global sensitivity analysis. Local sensitivity analysis methods are mostly based on partial derivatives of the output variables with respect to the input parameters of the model. They assume that the model is linear at least locally around the point of interest. These methods are computationally efficient and can be used to perform sensitivity analysis of models with a large number of parameters. However, these methods are not able to capture the non-linear effects of the parameters on the model output. Global sensitivity analysis methods, on the other hand, are based on the assumption that the model is non-linear. They generally involve sampling the parameter space and evaluating the model output at each sample point. Then the effect of the parameters on the model output is estimated using statistical methods such as variance decomposition or regression analysis. Given the inherent non-linearities and the number of parameters in agent-based models, global sensitivity analysis methods are more suitable for performing sensitivity analysis of these models. In this study, we use the Sobol method~\cite{Sobol2001} to perform a global sensitivity analysis of the model.

In this paper, we discuss the importance and necessity of sensitivity analysis of agent-based disease models and specifically study to what extent the geographical structure of the regions affects the parameter sensitivities to COVID spread. In other words, does it indeed matter that agent-based models explicitly resolves geographical aspects from maps, or should we use uniform geographies instead, saving ourselves development time and reducing simulation complexity? To this end, we use the Flu and Coronavirus Simulator (FACS)~\cite{Mahmood2022} to study the effect of the geographical distribution of the population on the spread of the disease. FACS is an open-source~\cite{Facs}, stochastic, spatially explicit and individual-based model that simulates the spread of an infectious disease in a population. The model simulates the spread of the disease by simulating the movement of individuals in the population and the transmission of the disease between individuals. The model has been used to study e.g. the effect of school closures on the spread of the disease~\cite{Conversation} and can be combined with a hospital model to support the allocation of intensive care capacity in anticipation of pandemic waves~\cite{Anagnostou:2022}.

We use FACS to simulate the spread of the disease in three different real-world regions: Călărași (Romania), Klaipėda (Lithuania), and Harrow (England). Each of these regions was selected by health authority stakeholders as part of the STAMINA research project~\cite{Stamina}, and has a different geographical structure in terms of the distribution of buildings and populations. Because sensitivity analysis procedures rely on a very large number of simulation runs, we use the SEAVEA toolkit~\cite{Seavea2023} to automatically deploy and run simulations on high-performance computers and analyze the results obtained. In our case, the simulations were run on the ARCHER2 supercomputer at EPCC in Edinburgh, UK, which has been used to support numerous large research projects around the world~\cite{Merchant2023,Payne2022,Bian2023}. Using these tools, we can automatically obtain the evolution of the disease as predicted by FACS and then compute the Sobol indices for a selected subset of input parameters.

In the Results section, we present the differences in the geographical structure of the regions and describe how they lead to differences in the movement patterns of individuals in the population and in the spread of disease. We also discuss the resulting variation in the spread of the disease in the three regions, present the results of the sensitivity analysis of the model and discuss how the observed differences in Sobol indices are related to differences in the geographical structure, movement of individuals and transmission of the disease in the three regions. In the Discussion section, we present the conclusions of the study and the implications of the results. We also discuss possible future research directions based on the results of this study. Details of the model and sensitivity analysis are provided in the Methods section. This section also details the computational resources used for the study.



\section*{Results}

In this section, we present the various aspects of the results obtained by simulating the spread of an infectious disease in the regions of Călărași, Klaipėda, and Harrow using FACS. Although the disease used for this study is the COVID-19 virus, the results are applicable to any infectious disease. The regions of Călărași, Klaipėda and Harrow have been chosen because, in spite of having similar populations, they have significantly different geographical structures in terms of the distribution of houses and amenities.


\subsection*{Geographical structure of the regions}

\begin{figure}

    \centering
    \includegraphics[width=0.8\linewidth]{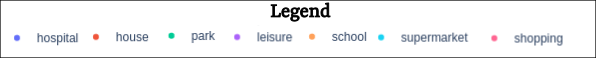}\\~\\

      \begin{subfigure}{0.33\textwidth}
        \includegraphics[width=\linewidth]{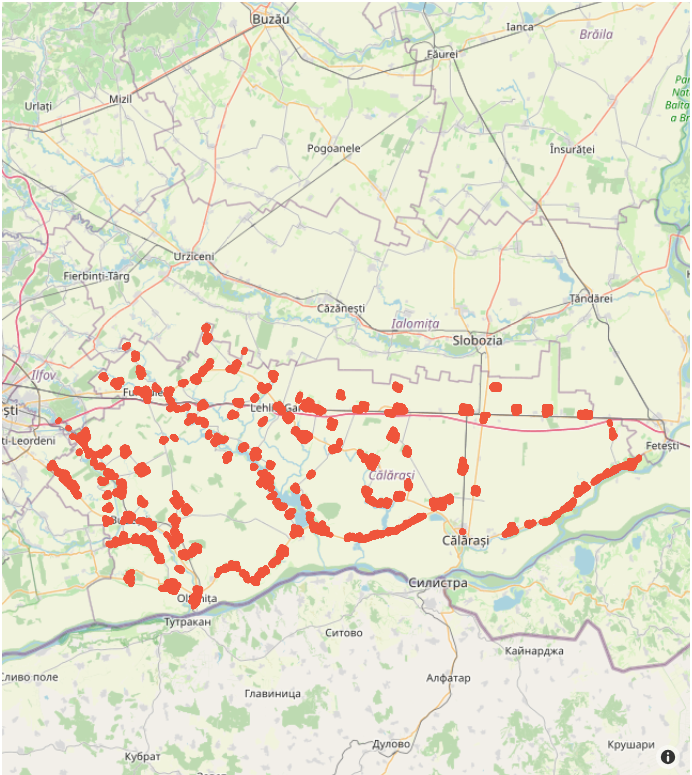}
        \label{fig:calarasi1}
      \end{subfigure}
      \begin{subfigure}{0.33\textwidth}
        \includegraphics[width=\linewidth]{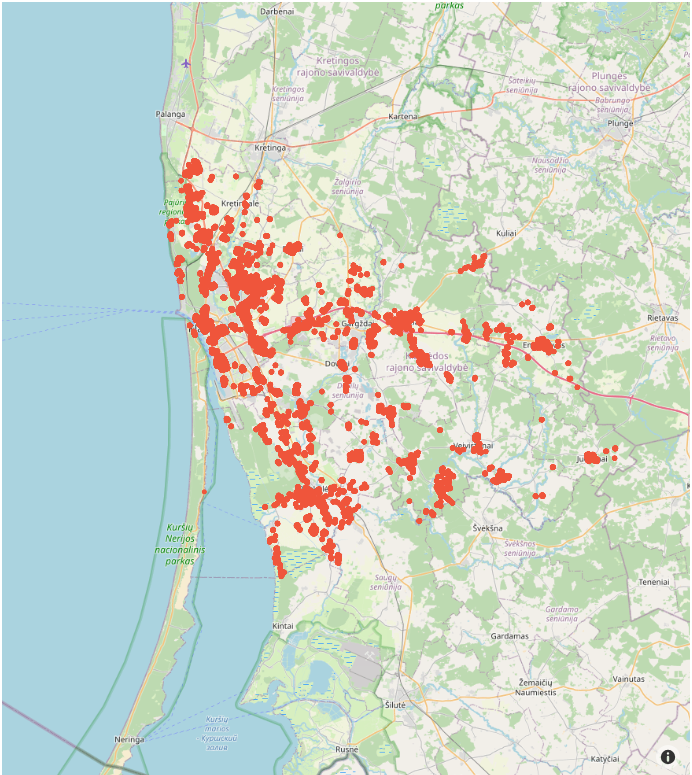}
        \label{fig:calarasi2}
      \end{subfigure}
      \begin{subfigure}{0.33\textwidth}
        \includegraphics[width=\linewidth]{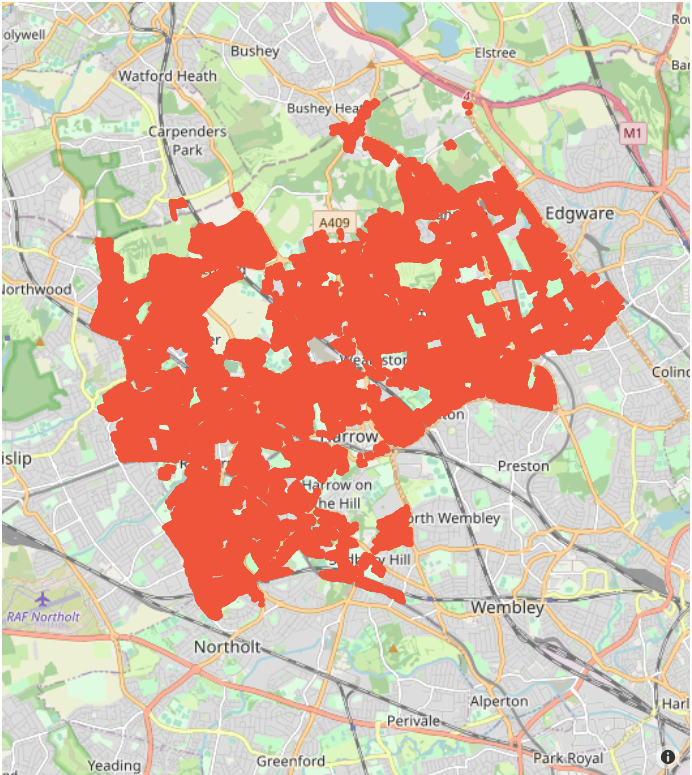}
        \label{fig:klaipeda1}
      \end{subfigure}
      \begin{subfigure}{0.33\textwidth}
        \includegraphics[width=\linewidth]{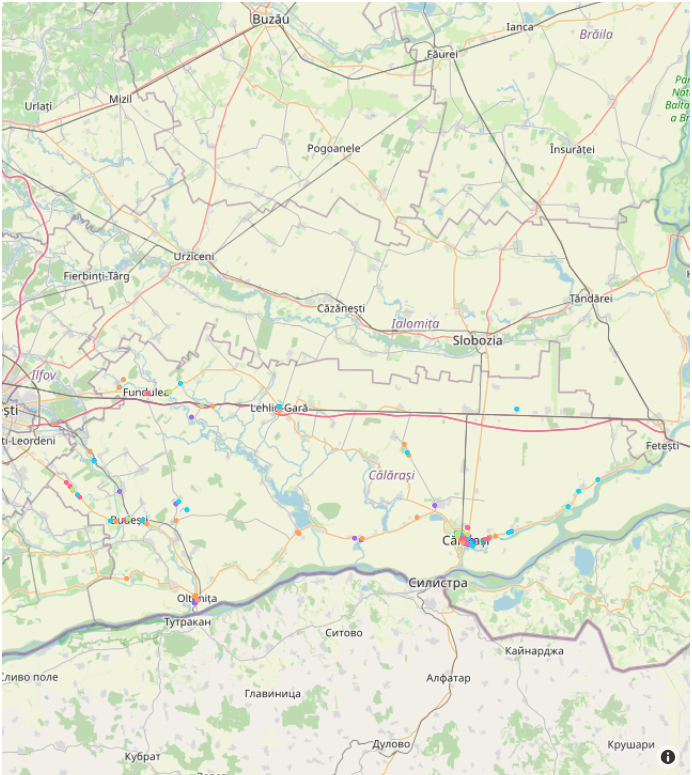}
        \caption{Călărași}
        \label{fig:klaipeda2}
      \end{subfigure}
      \begin{subfigure}{0.33\textwidth}
        \includegraphics[width=\linewidth]{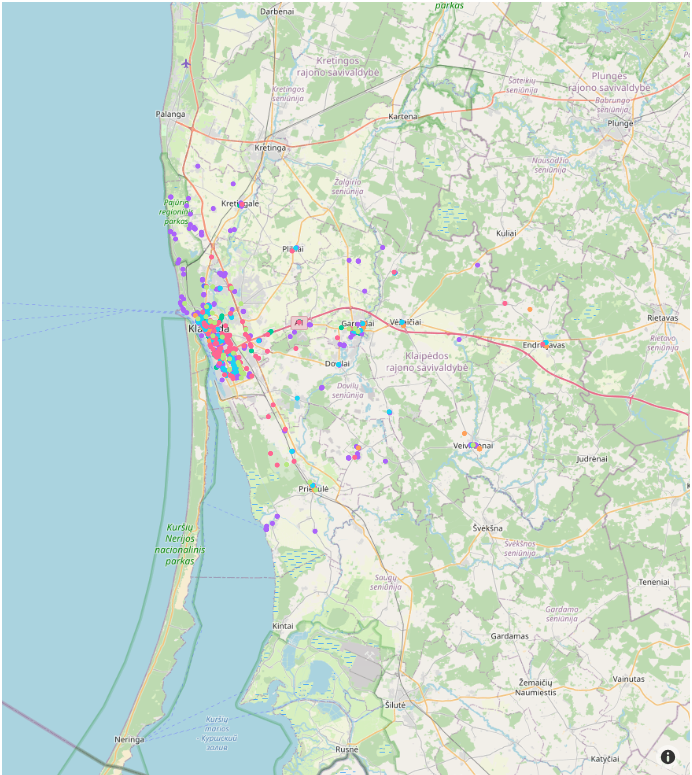}
        \caption{Klaipėda}
        \label{fig:harrow1}
      \end{subfigure}
      \begin{subfigure}{0.33\textwidth}
        \includegraphics[width=\linewidth]{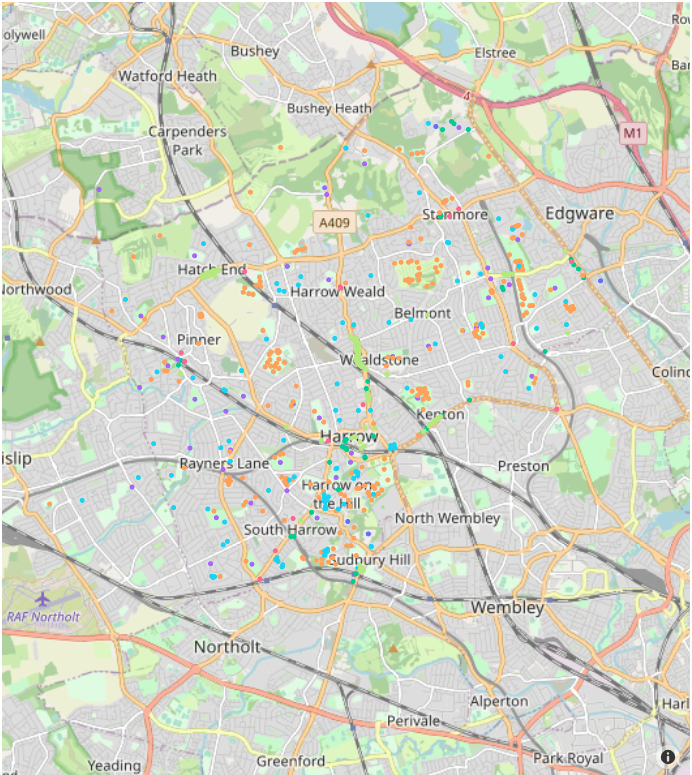}
        \caption{Harrow}
        \label{fig:harrow2}
      \end{subfigure}
    
    \caption{Maps of (a) Călărași, (b) Klaipėda, and (c) Harrow. Maps in the top and bottom rows of each column show the location of houses and amenities in each region respectively, as identified from OpenStreetMap. While amenities are individually identified, houses are randomly generated in the housing areas identified from OpenStreetMap. Offices are not shown here as they are assumed to be uniformly generated throughout the region.}
    \label{fig:maps}

\end{figure}

To appreciate further results, we first present the distribution of buildings in Călărași, Klaipėda, and Harrow (Figure~\ref{fig:maps}) as identified from OpenStreetMap using the method described in the Methods section. Note that the offices in the regions have not been shown because they are uniformly distributed across the entire region (see the Methods section). The maps make it evident that the houses and other amenities, namely hospitals, parks, leisure centers, schools, supermarkets, and shopping centers, are distributed very differently across the regions.

In Călărași, there are distinct clusters of houses in the suburbs and a single urban area in the southeast of the region where most of the amenities are located. This urban area has relatively few houses, which implies that most of the population travels to a relatively small urban center to use the amenities.

In Klaipėda, the distribution of houses and amenities is around multiple population hubs. Each of these hubs has houses as well as amenities. The population of each hub would mostly use one of the  amenities present within the hub and would rarely interact with populations from other hubs during their visits to the amenities.

In Harrow, the population is very densely distributed with a relatively homogeneous distribution of amenities across the region. This allows for a relatively even mixing of the population as people visit the amenities in the region.

We are aware that the methods employed for extracting the buildings from the regions might not be completely accurate due to inaccuracies in the OpenStreetMap data, its varying accuracy in different parts of the world, as well as the inaccurate choice of classification criteria in our algorithm. However, it is important to note that in this study, we are not interested in the actual prediction of the evolution of a disease in these cities. Instead, we focus on the qualitative difference in the structure of the city and the effect it has on the evolution of an infectious disease. Therefore, we are not concerned about the accurate representation of the regions in particular but the relative distribution of houses and amenities in the region.

\subsection*{Location graphs}

Based on the geographical locations of the buildings, a bipartite location graph of buildings is created for each region. Each house in the region is connected to a single amenity of each type. This leads to each house being connected to seven amenities. The choice of amenities to be connected to each house depends on the physical size of the amenity and its distance to the house (refer to the Methods section for further details). Therefore, while the degree of each house is identical, the mean degree of each amenity depends on its geographical location. Note further that the degree distribution of offices is similar for each region because of the assumed uniform geographical distribution of offices in the regions.

\begin{figure}
    \includegraphics[width=\textwidth]{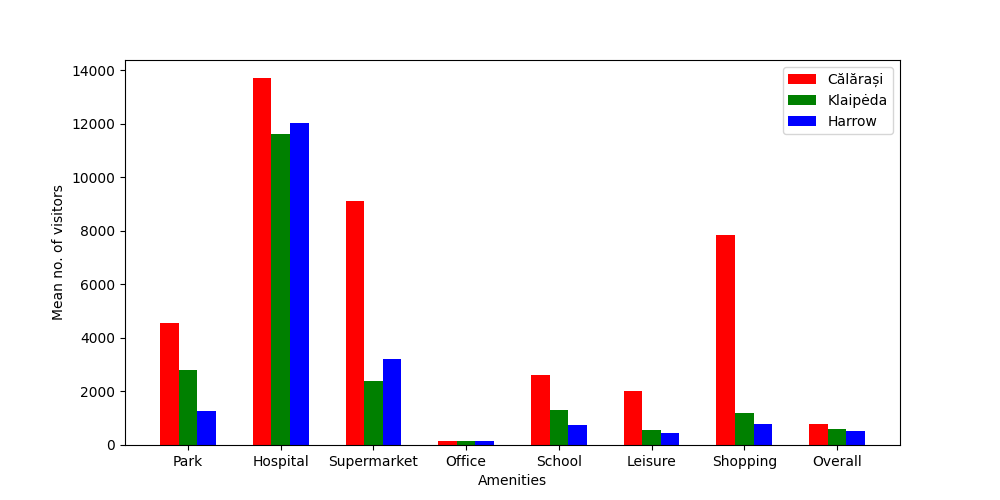}
    \caption{Mean degree of each type of amenities in the three regions. Since amenities are connected to the houses which use the amenity, the degree of an amenity indicates the number of people who visit the amenity. Note that the degree of an amenity is independent of the actual time spent by a person in an amenity, which depends on the type of the amenity and the age of the person.}
    \label{fig:degree}
\end{figure}


The above-mentioned difference in the physical distribution of houses and amenities is reflected in the structure of the location graphs of Călărași, Klaipėda and Harrow. Figure~\ref{fig:degree} shows the mean degree for nodes representing each type of amenity for each region. Since the amenities are fewer and more centrally located, the mean degree for each type of amenity is the highest in Călărași. This is followed by Klaipėda and Harrow, in order, for all amenities except hospitals and supermarkets. This is a further indication of the difference in the geographical structures of regions. The degree of each amenity represents the number of houses served by the amenity. Since the household size is taken to be the same across regions, the differences in mean degrees also represent the number of people visiting these amenities on average. In the next subsections, we would see the impact of this difference on the spread of the disease in these cities.

\subsection*{Disease progression}

\begin{figure}
    \centering
    \includegraphics[width=\textwidth]{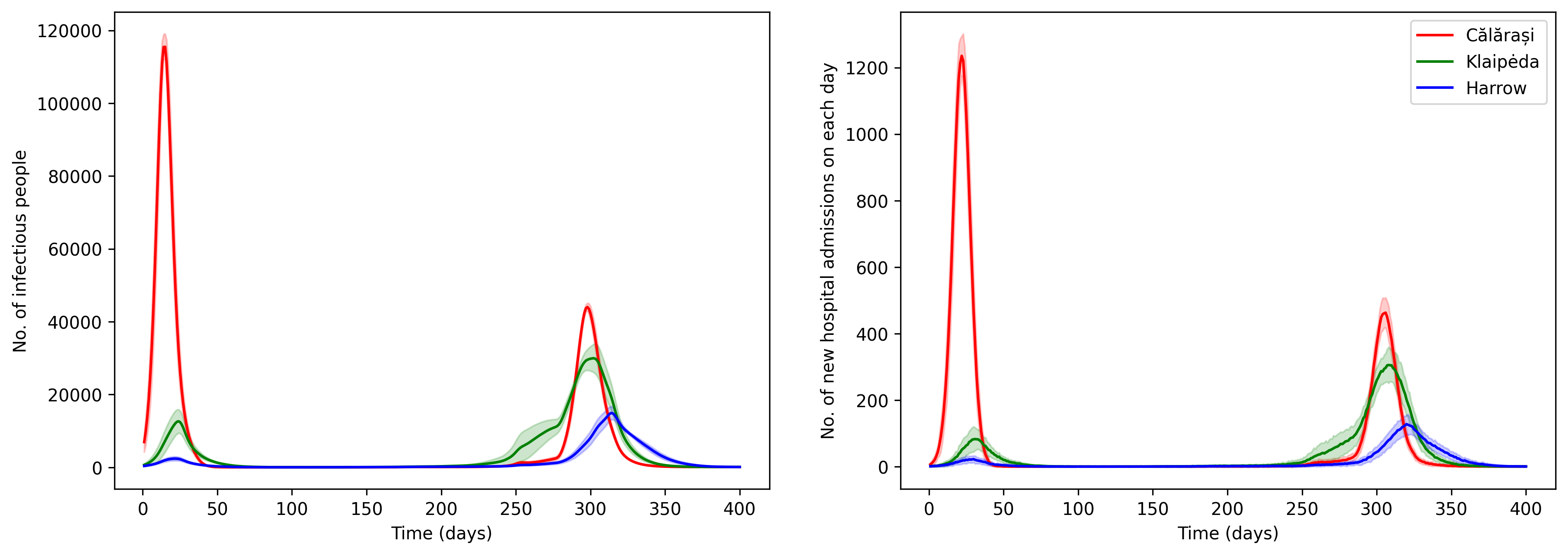}
    \caption{Number of infectious people and the number of hospitalizations on each day of simulation for Călărași, Klaipėda and Harrow. Since the simulations are stochastic in nature, the solid lines in the plots show the average of the 50 simulation results. The 95\% confidence intervals are shown using shaded regions. The lockdown measures are kept constant across all runs.}
    \label{fig:timeline}
\end{figure}

We now present the results of the FACS simulation. To analyze the differences in sensitivity analysis caused by differences in the spatial features of the region, we simulate the progression of the disease from March 1, 2020, for a duration of 400 days in the regions of Călărași, Klaipėda, and Harrow, keeping all other parameters identical. In particular, the demographics of the regions, properties of the disease, needs of the individual agents, government measures and lockdowns, and vaccination strategies are kept identical across the regions. The details of the parameters of simulations can be seen in the Methods section. Given the stochastic nature of the simulations, an ensemble of 50 simulations was run for each region, and the mean of the results along with their 95\% confidence intervals is plotted in Figure~\ref{fig:timeline}.

In Figure, we show the number of infectious people and the daily number of new hospitalizations as a function of time. It is evident from the plots that all regions witness two waves of infections during the period of simulation. However, the shape and height of the waves differ across the regions. It is also to be noted that the plot for the number of hospitalizations is similar to the number of infectious people in shape but is scaled down in magnitude. This is expected, as only a fraction of infected people need to be hospitalized.

In the region of Călărași, a higher level of intermixing among individuals is observed, mainly due to the increased number of visitors that frequent the local amenities. As a result, the peaks in the number of infections and hospitalization are higher and exhibit a sharper intensity compared to other regions. On the contrary, in the areas of Klaipėda and Harrow, the heights of these peaks are due to a lower number of individuals visiting the same amenities.

Another important peculiarity in the progression of the disease in Klaipėda is the distinct shape of the second wave. Around day 250 after the start of the simulation, the number of infectious people, as well as the number of hospitalizations, start increasing gradually. The simulation then reaches an inflection point around day 275, where the rate of increase of infectious people and hospital admissions suddenly increases. Such an inflection point is not found in the other two regions. We analyze the reasons behind this peculiarity in the next subsection.

\begin{figure}
    \centering
     \makebox[\textwidth][c]{\includegraphics[height=0.38\textwidth]{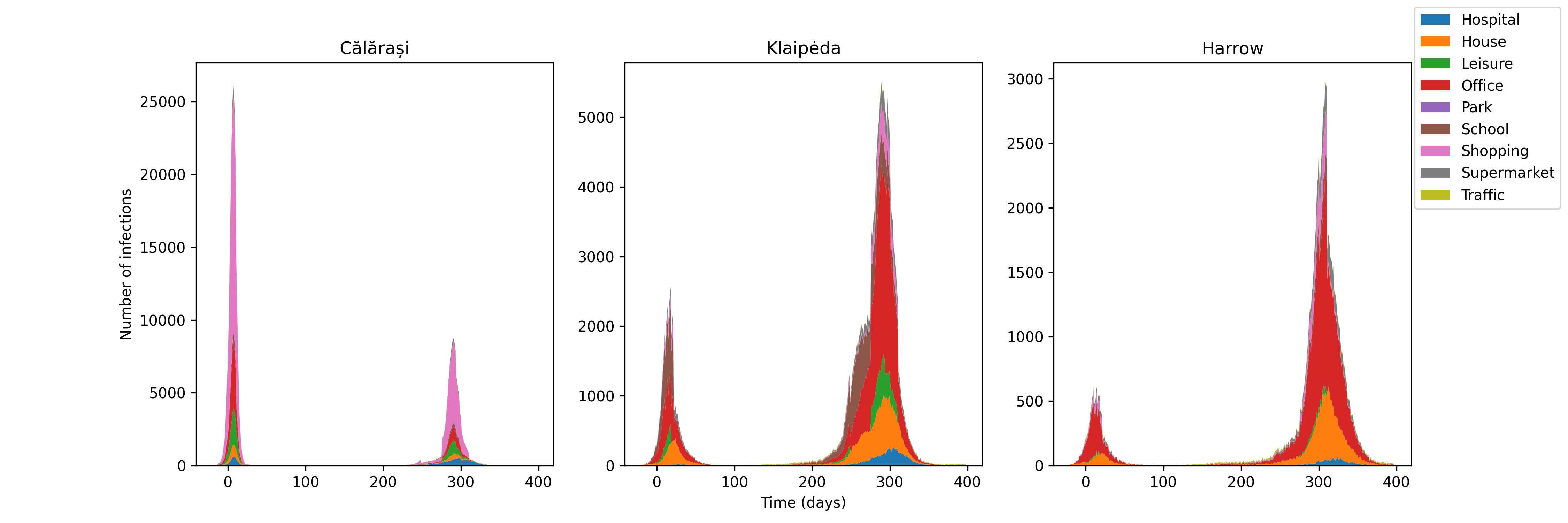}}
    \caption{Stacked plot showing the number of infections occurred in each type of amenity on each day in Călărași, Klaipėda and Harrow. On each day, the number of infections occurred at each type of location is represented by a different color.}
    \label{fig:stacked}
\end{figure}

\begin{table}[ht]
    \centering
    \caption{Summary of lockdown measures implemented during the simulation.}
    \begin{tabular}{|c|c|c|c|}
        \hline
        Date               & Days since start of simulation & Lockdown Measures  \\
        \hline \hline
        March 01, 2020     & 0                              & Start of simulation       \\
        March 21, 2020     & 20                             & First lockdown            \\
        April 22, 2020     & 52                             & Peak of lockdown measures \\
        September 01, 2020 & 184                            & Schools re-opening   \\
        November 05, 2020  & 249                            & Second lockdown          \\
        December 02, 2020  & 276                            & Restrictions lifted        \\
        December 23, 2020  & 297                            & Christmas bubble         \\
        January 06, 2021   & 311                            & Third lockdown            \\
        March 08, 2021     & 372                            & Restrictions lifting      \\
        April 04, 2021     & 399                            & End of simulation         \\
        \hline
    \end{tabular}
    \label{tab:measures}
\end{table}

One of the crucial factors affecting the overall shape of the plots shown in Figure~\ref{fig:timeline} are the lockdown measures taken during the simulation period. These measures are summarized in Table~\ref{tab:measures}. Prior to the start of simulation, the population is randomly seeded with a fixed number of initial infections. Subsequently, based on the movement and interaction of people, the number of infections changes over time. The number of infections that occur each day is plotted in Figure~\ref{fig:stacked}. For each day, the number of infections occurring at each location type is shown in different colors.

It is important to note that the timeline presented in Table~\ref{tab:measures} only shows the dates on which the main policy decisions were made. The FACS simulation used for the results presented here takes into consideration many more minor measures. However, only some of the major measures are shown in the table due to their relevance in the results presented here. A full list of all measures used for the simulations can be found on the GitHub page of the software~\cite{config_files_1,config_files_2,config_files_3}.

Looking at Figures~\ref{fig:timeline} and \ref{fig:stacked} together with Table~\ref{tab:measures} gives us a clear idea about how the lockdown measures have affected the evolution of the disease in the regions being studied. Since the first lockdown measures come into effect only 20 days after the start of the simulations, the the number of infections increase rapidly from the start of the simulation to around day 20. This causes the first wave of infections and hospitalizations in the regions. From day 20 to day 53, various measures are introduced so that the interaction of people and the transmission of diseases is restricted. This includes allowing work from home when possible, partial and then complete closure of schools and leisure centers, mask mandates, travel restrictions and emphasis on social distancing.

These measures lead to the end of the first wave in all the three regions by day 75. Thereafter, the lockdown measures are gradually lifted from around day 184. It is on this day that schools start re-opening gradually with a varying but limited fraction of students being allowed to attend. Other restrictions are also gradually lifted which results in a lesser fraction of population working from home and a greater fraction of population intermingling in shopping centers, leisure centers and parks. The effects of lesser restrictions are not directly visible until after day 200 when the number of infectious people and hospitalizations start increasing again.

This rise in the number of infections after day 200 gives rise to the second wave of infections and hospitalizations in the three regions. While the heights of the peaks of infection and hospitalization follow the same trend as the first wave, the shape of the second wave are significantly different for the three regions being studied. The second wave is sharp and symmetric around the peak for Călărași, but for Harrow, the wave is asymmetric with a long tail. As also noted earlier, the second wave in Klaipėda is characterized by an inflection point around day 275 after which the slope of the two curves in Figure~\ref{fig:timeline} increase sharply. We now discuss these differences and the possible reasons behind them.

\subsection*{Comparing results of the regions}

Note that there is a clear distinction among the regions when it comes to the locations at which the infections take place (see Figure~\ref{fig:stacked}). While in Călărași most of the infections occur in the shopping centers, the majority of infections in Klaipėda and Harrow occur in houses and offices. Additionally, in Klaipėda, a significant share of the infections also occur in schools. These differences in the location of infections are a direct result of differences in the structure of the location graphs and the chronology of the lockdown measures.

At the start of the simulation, the number of infections increases with time as people interact with each other. In Călărași, due to a high degree of connectivity for all the amenities (except offices), more people visit the amenities at the same time. This leads to a quick spread of infections resulting in a sharp and high peak for the first wave. The wave comes down in Călărași by day 50 mostly because of the lockdown restrictions as well as due to the high levels of immunity developed across the population. The high immunity in the population is perhaps also one of the reason behind the late start of the second wave as compared to the other regions. Although schools and other amenities start opening up gradually, it is not until the restrictions are significantly reduced after day 276 that the second wave starts picking up. As evident from Figure~\ref{fig:stacked}, most of the infections in Călărași happen in shopping centers. This is because of the high connectivity of shopping centers (as seen in Figure~\ref{fig:degree}). It is important to note that although the average degree of hospitals and supermarkets are higher than shopping centers in Călărași, they do not have a large contribution towards infections. This is because hospitals and supermarkets typically have a larger area than shopping centers. This allows people to spread out more, resulting in a lesser probability of infections. Infections in hospitals are also less due to the less amount of time that people spend in them on average and the additional isolation measures found in them.

In Klaipėda and Harrow, the average degree of an amenity location is lower than that of Călărași. This leads to less number of people visiting any particular amenity on average and therefore, a lower height of waves. In particular, this leads to a late peak of the first wave. It is also to be noted that due to a lower peak of the first wave, a significantly lower number of people got immunized. This leads to a second wave which is higher than the first wave. It is also interesting that most of the infections in Klaipėda and Harrow occur in houses and offices (see Figure~\ref{fig:stacked}). This is due to the significantly larger portions of the day spent in them. At least a small but significant portion of the population continues working in offices throughout parts of the simulation period. This causes infections to spread first in offices and then across houses where they spent the majority of their time.

Schools also play a significant role in spreading infections in Klaipėda due to their high degree in the location graph. This leads to the interesting infection point in the second wave as noted earlier. After the first wave, schools reopen on day 184. This leads to an increase in infections in schools. The infected students from schools then spread the infection to other people in the household. They spread the infection further mainly through the offices. The spread of infections is accelerated when movement restrictions are lifted between days 276 and 297. In particular, shopping centers and supermarkets follow normal opening hours, movement between regions is permitted again and work from home is significantly reduced. This leads to the inflection point noted earlier.

Due to a lower average degree of schools in Harrow, a significantly lower number of infections take place in schools. Therefore, a significant upsurge in the number of infections is not seen for Harrow after the re-opening of schools on day 184. Hence, the onset of the second wave is delayed by approximately 75 days when compared to Klaipėda. Other than that, the overall lower heights of the waves in Harrow can be attributed to the lower average degree of amenities.

\begin{figure}
    \centering
     \makebox[\textwidth][c]{\includegraphics[width=1.25\textwidth]{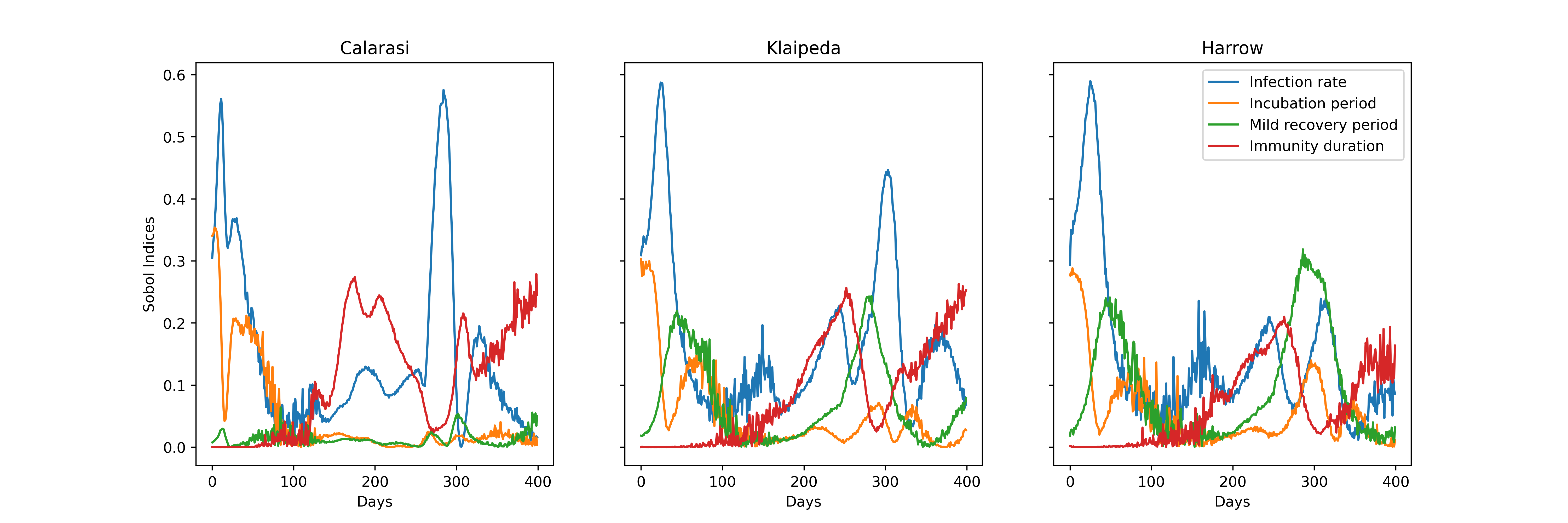}}
    \caption{Sobol indices corresponding to four disease parameters as a function of time for Călărași, Klaipėda and Harrow with respect to the number of hospital admissions on each day.}
    \label{fig:sobol}
\end{figure}

\subsection*{Sobol indices}

The FACS simulation depends on a large number of parameters. Location and size of buildings, the age distribution of the population, time spent by each agent in each amenity, restrictions imposed on the population during the lockdown, specific properties of the disease vector, and the efficiency and rates of vaccines administered are some of the broad categories of properties on which the results of the simulation depend. Each of these parameters affects different aspects of the simulation in different ways. To study the impact of some of these parameters, we present the sensitivity analysis of the results presented earlier with respect to some of the simulation parameters.

With the devastation caused around the world by new coronavirus mutants in the COVID-19 pandemic, we chose to analyze in detail the importance of the properties of the disease vector. This would allow us to address the following class of questions: If a new mutant of the disease vector arises, which of its properties would impact the number of hospitalizations the most? In FACS, the disease vector is characterized by seven scalar properties: (a) infection rate, (b) mortality period, (c) recovery period, (d) mild recovery period, (e) incubation period, (f) period to hospitalization, and (g) immunity duration. Among these parameters, our preliminary analysis showed that the Sobol indices corresponding to the mortality period, recovery period, and period to hospitalization were significantly lower than those corresponding to the other four parameters. Therefore, given the large size of the parameter set, we chose a smaller subset for our detailed analysis. It may be noted that the chosen subset of parameters also overlaps with one used in a small-scale sensitivity analysis in an earlier publication~\cite{Groen2021}.

In Figure~\ref{fig:sobol}, we present the results of the sensitivity analysis with respect to four scalar parameters which determine the properties of the disease vector: (a) infection rate, (b) incubation period, (c) mild recovery period and (d) duration of immunity. The plot, for each region, gives the first-order Sobol indices corresponding to each of these parameters as a function of time. The Sobol index is a number from 0 to 1 which quantifies the sensitivity of the simulation result with respect to these parameters. Since the output of FACS is multi-dimensional, we select the number of daily hospitalization to be the output with respect to which the Sobol indices are to be computed.

Although the computation of Sobol indices is a computationally intensive process, its interpretation is relatively simple to explain. If the Sobol index corresponding to a parameter is high, the output of the simulation is more sensitive to that parameter. In other words, relatively small variations in a parameter with high Sobol index will lead to relatively high variation in the output of the simulation and vice versa.

In order to explain the plots observed in Figure~\ref{fig:sobol}, let us take a closer look at the parameters with respect to which the Sobol indices are computed.

\subsubsection*{Infection rate}

Infection rate is the probability that a susceptible individual gets infected when he/she remains in the proximity of an infected individual for 24 hours. Since it is necessary for an individual to be in the infected state before he/she can get hospitalized, the infection rate of the disease directly impacts the daily number of hospitalizations. However, note that only the susceptible individuals can be infected and not those who are immune. Moreover, infections can only spread if individuals interact with each other. Therefore, the Sobol index of infection rate increases when the number of susceptible individuals is high and the lockdown restrictions allow a sufficient number of interactions between susceptible and infected individuals.

This trend is clearly seen in Figure~\ref{fig:sobol}. In the initial periods of simulation, the Sobol index corresponding to infection rate starts high for all three regions. It drops rapidly as the first wave ends and the number of infections comes close to zero. This is due to the increased number of immune individuals as well as the imposition of lockdown restrictions. As minor modifications are regularly made to the lockdown measures between the two waves, the Sobol index corresponding to the infection rate fluctuates around a low value. As the restrictions are gradually lifted and the number of infections gradually increases, the Sobol index also increases. The curve corresponding to the infection rate shows consecutive peaks until around day 300 when the second wave is seen in all regions.

While the overall trend of the Sobol index is the same across regions, there are slight differences between the three regions. In Călărași, the peak Sobol index for infection rate is similar for the first and second waves. However, the peak Sobol index for infection rate for the second wave is smaller than that of the first wave for Klaipėda and Harrow. A lower peak for the second wave in Klaipėda and Harrow implies that the daily number of hospital admissions is also determined by several competing factors other than infection rate. However, in Călărași, the infection rate is the primary and overwhelming factor determining the number of daily hospitalizations during the second wave. We will now look into this aspect of the mechanism as we analyze the Sobol indices corresponding to other parameters.

\subsubsection*{Incubation period}

When a susceptible individual gets exposed to the disease vector, there is a small period of time before the symptoms of the disease manifest themselves.  This period of time is called the incubation period. Since the incubation period essentially delays the time when symptoms are perceived and hence the individual is potentially hospitalized, the impact of changes in the incubation period is high if the number of hospitalizations changes in a short interval of time. Hence, in such a period, the Sobol index corresponding to it is high.

In Figure~\ref{fig:sobol}, we can clearly see that Sobol indices corresponding to incubation period starts high for all regions because, during the initial days of simulation, the number of infections and hospitalizations rise rapidly.  The Sobol indices then start falling until after the end of the first wave in all regions.  They stay low during most of the first lockdown period rising only just before the begining of the second wave. The variation in time, duration and shape of the second wave across the regions is captured by the variations in Sobol indices of incubation period. In Călărași, where the second wave is sharp, the Sobol index shows slight bumps just before and after the start of the wave when the number of infections and hospitalizations changes the fastest. In Harrow, the rise and fall of the Sobol index is much more pronounced and prolonged corresponding to the long-lasting second wave in the region. Interestingly, the Sobol index for Klaipėda captures the peculiar shape of the second wave. The Sobol index rises not only prior to the advent and after the end of the second wave but also during the infection point in the wave around day 300.

It is notable that, while changes in the rates of hospitalization are captured well by the Sobol indices corresponding to them, incubation period by itself does not become a parameter to which the results of the simulation are most sensitive. This might be due to the fact that neither the peak height  nor the frequency of the waves of infection and hospitalization can be significantly altered by independently changing the incubation period of the disease. Changing the incubation period can only alter the time when the waves occur.

\subsubsection*{Mild recovery period}

Mild recovery period is the average number of days required for recovery if an infected person is showing mild symptoms. An infection is considered mild if the infected person does not need to be hospitalized. In FACS, there are two ways in which an infected person can get hospitalized. After infection, the person might develop severe infection immediately. In this case, he/she is directly admitted to the hospital and is never isolated at home. However, If the symptoms are mild, he/she would stay isolated at home.  Until such a person recovers, there is a probability that the disease may worsen and he/she might be hospitalized. It is in this scenario that mild recovery period becomes crucial in affecting the overall number of hospitalizations per day.

The stark differences between the sensitivity of the daily number of hospitalizations on mild recovery period for the three regions can be seen clearly in Figure~\ref{fig:sobol}. In Călărași, where the waves of infection are sharp and high, changes in mild recovery period have a very limited impact on the number of hospitalizations. In Klaipėda and Harrow, where the waves are less sharp and high, the impact of mild recovery period is more pronounced.

This is probably because, when the waves of infection are sharp, people remain infected for a short period of time. Since mild recovery period can affect the daily number of hospitalizations only when the number of infected people is significant, therefore, there is only a short interval of time when it can have a high Sobol index. Additionally, it should be noted that a high peak implies a large outbreak in a specific set of infections in a small interval of time. This results in a large number of infected people becoming hospitalized in a small period of time. This results in a sudden reduction in the number of people who can spread the infection further. Therefore, the impact of mild recovery period would be limited. This explains why mild recovery period has a very limited impact in Călărași.

In Klaipėda and Harrow, the waves of infection are extended over a longer period of time. This allows for more opportunities for interaction between infected and susceptible people resulting in more infected people over time. The mild recovery period parameter impacts the hospitalization probability of these newly infected people. The higher the mild recovery period, the greater the uncertainty about the time at which the person might be hospitalized. This effect is seen most prominently during the second wave in Harrow which takes a long time to subside. This results in a significantly high Sobol index for mild recovery period during the second wave.

\subsubsection*{Immunity Duration}

Immunity duration is the average number of days for which a person remains immune from the disease after gaining immunity. This immunity can be gained due as a result of recovery from a recent infection or vaccination. In the simulation results shown here, vaccination was gradually introduced from day 275.  Prior to that, all immunity in the population is induced as a result of prior infections.

Since the duration of immunity directly impacts the number of people who can become infected, it plays a significant role in determining the daily number of hospitalizations in all regions studied. In Călărași, where a significant proportion of the population was infected during the second wave, the Sobol index corresponding to immunity duration is observed for the longest duration. In Klaipėda and Harrow, the impact of immunity duration is significantly observed only for a shorter and later period of time. This difference can be attributed to the lower number of people infected in these two regions during the first wave.

In Călărași, a sharp and high first wave results in a large number of people recovering and gaining immunity in a short period of time. Therefore, a change in the duration of immunity between the first and second waves would change the time when a large number of people would become susceptible again. This would change their probability of getting infected and hence getting hospitalized. In Klaipėda and Harrow, the first wave is much wider and shorter. Therefore, the number of people who became immune in the first place is smaller. Moreover, the period of time, when they gain immunity, is spread over a longer period of time. Therefore, the Sobol index in these regions gradually gains importance over time until the advent of the second wave.

After day 275, when gradual vaccination was introduced in the population, the Sobol index corresponding to immunity duration increases for all regions due to vaccine-induced immunity.

\section*{Discussion}

Through a detailed study of simulation results of the agent-based model FACS in three geographically distinct regions of Europe, we have demonstrated that geographical structure of regions has a significant impact on the role of each input parameter in the model. The three regions studied in this paper: Călărași, Klaipėda and Harrow have similar population sizes but very different geographical distributions of houses and amenities. We have shown that this difference leads to differences in population movement patterns, evolution of the disease and sensitivities to disease parameters.

While all three regions have experienced two waves of infection, the timing and intensity of the waves have been different. Călărași, with a more segregated population, witnessed the sharpest and most intense waves, with a large number of infections and hospitalizations occurring in a short period of time. Klaipėda and Harrow, which have relatively more mixed populations, saw waves which were flatter but lasted for a longer period of time. Some of the intricate details of the shapes and timing of the waves were discussed in detail in the previous sections. We also highlight the relationship between differences in the geographical structure of the regions and differences in the evolution of the disease.

We also analyzed the Sobol indices of a subset of parameters in the FACS model to highlight that the differences in the geographical structure of the regions also impact the sensitivity of the model to different parameters. One of the interesting observations was that, for instance, the recovery period from a mild infection of the disease has a very limited impact on the number of hospitalizations in a region like Călărași, where infections spread through a limited number of highly connected hubs. However, in regions like Klaipėda and Harrow, where amenities are more evenly distributed, the recovery period from mild infection has a more pronounced impact on the number of hospitalizations.

The results of the sensitivity analysis presented here give us an insight into the impact of the various properties of the disease virus on different types of geographical regions. Given the various mutations that may arise in the genetic structure of the virus, it is important to understand the impact of these mutations on infections and hospitalizations in a region. The results presented here highlight that changes in the same property of the virus may have different impacts on the evolution of the disease in different regions. Therefore, it is important to understand the geographical structure of a region to make accurate predictions about the evolution of the disease in that region.

Due to the large number of parameters in the FACS model, it is not possible to analyze the Sobol indices of all parameters in this paper. However, the results presented in this paper highlight the importance of analyzing the sensitivity of the model to different parameters. This can help identify parameters that have a significant impact on the evolution of the disease in a region. This can also help identify the parameters that need to be estimated more accurately to make better predictions in a region. For example, the impact of various lockdown measures on the evolution of infections and hospitalizations is also an important factor that needs to be analyzed. The effectiveness of these measures will also depend on the geographical structure of the region. In particular, the results presented in this paper show that infections in schools play a significant role in Klaipėda. Therefore, school closure periods are expected to have a greater impact on the evolution of the disease in Klaipėda as compared to Călărași and Harrow. Similar statements can be made about shopping centres in Călărași and offices in Harrow. Since lockdowns also have an economic impact, it becomes interesting and important to conduct such sensitivity analysis in each region to mitigate any future pandemics in a more effective manner.

\section*{Methods}
\label{sec:Methods}


At the core of the results discussed in the previous section, are the Flu and Coronavirus Simulator, computation of Sobol indices, and the software and hardware architecture which allows such computationally intensive tasks to be performed. In this section, we present each of these aspects in detail.

\subsection*{Overview of Flu and Coronavirus Simulator (FACS)}

All the results presented in this paper were generated by the Flu and Coronavirus Simulator (FACS) which simulates the propagation of an infectious disease in a given geographical region. FACS views the geographical region as a set of spatially distributed houses and amenities. Amenities are classified into seven categories: (i) shops, (ii) supermarkets (iii) schools, (iv) offices, (v) parks, (vi) leisure, and (vii) hospitals. Each house contains one or more agents (persons). Some of these agents are initially infected by the disease being simulated. FACS simulates each day for all the agents who visit these amenities according to their age and spend some time of the day there. During these visits, the agents might either get infected or infect other people with the virus.

In terms of the disease model, the agents are divided into the following categories: susceptible, exposed, infected, recovered, dead, and immune. Depending on the visits made by the agent during the day, and other factors such as the lockdown measures and vaccination status, the transition probabilities among the categories are computed. For this computation, the age of the agent, the size and type of the locations visited, and the compliance rate to the lockdown measures is also considered. The transition rates can also be modified with the progress of time. This is crucial to model diseases where new variants of the disease vector emerge over time.


The results of a particular run of FACS are defined by six configuration files:
\begin{enumerate}
    \item \textbf{Buildings file:} lists the geographical coordinates of houses and other amenities in the geographical region. For each amenity, its size in terms of the area occupied on the map is also listed.

    \item \textbf{Demographics file:} describes the age-dependent population of the region

    \item \textbf{Needs file:} defines the amount of time spent by each agent at each type of amenity location. This depends on the age of the agent.

    \item \textbf{Disease file:} defines the properties of the disease vector such as its infection rate, the probability to get hospitalized, incubation period, etc.

    \item \textbf{Measures file:} gives a timeline of the social and governmental measures being taken to mitigate the spread of the disease such as movement restrictions, mask mandates, restricted opening hours of certain amenities, etc.

    \item \textbf{Vaccinations file:} defines the time-dependent rate of vaccination and vaccine efficiency.
\end{enumerate}

Once all configuration files are provided, FACS can be run using a one-line command on any Linux-based system. Some of the initial conditions for the simulation, some parameters, and the input/output directories are specified as options in the command. The runtime of FACS depends on the population size of the geographical region, parameters and initial conditions.

As a first step, FACS constructs a location graph of the region based on the locations provided in the buildings file. This involves connecting each house of the region with one amenity of each type. This is done by computing a cost-function $C_{ij}$ for each house $i$ and amenity $j$,

\begin{align}
    C_{ij} = \frac{D_{ij}}{\sqrt{S_j}}
\end{align}

where $D_{ij}$ is the distance between the house and the amenity, and $S_j$ is the size of the amenity. Then, for each ammenity category, a link is made between house $i$ and amenity $k$ such that $C_{ik} = \min C_{ij} \forall j$.

Thereafter, each house is randomly populated by one or more agents. Each agent is characterized by an age which is sampled in accordance with the demographics file.

Thereafter, FACS simulates the daily routine of each agent from a specified date. The agents, based on their age, try to spend a certain amount of time in a day at each amenity to which their house is connected. Whether they can visit the amenity successfully on a particular day depends on the restrictions applied on that day, as well as their infection status. For example, if the agent is hospitalized, they neglect their needs and spend their entire day at the hospital. During these visits to the amenities, the agents may get exposed to the disease vector via other infected people. They may then pass through the various stages of infection as described earlier.


As its main output, FACS computes timeseries of the various quantities of interests such as the number of susceptible, exposed, infectious people, as well as the number of hospitalizations and deaths. In addition, the simulator produces output files detailing the locations with infections and recoveries that occurred on each day.

\subsubsection*{A note on the generation of offices and in FACS}

Offices in FACS represent all types of workplaces in the region. Although the locations of houses and other types of amenities used in FACS are in accordance with their locations in OpenStreetMaps, the office buildings used in FACS simulation are assumed to be uniformly distributed across the regions in consideration. This is primarily because a significant part of the population travels to other neighboring regions for work. Additionally, given the diversity of types of workplaces, it was particularly difficult to create an exhaustive list of tags that might be used to identify buildings as workplaces. With these shortcomings in the available data, runs conducted for preliminary validation studies demonstrated that the model is better validated with offices being uniformly distributed across the region.

\subsection*{Sobol indices}

The sensitivity analysis presented in this paper primarily involves the computation of Sobol indices. We now discuss in brief the computation of these Sobol indices.

Let there be a time-varying function $y = f(x; p): \mathcal{R}^n \rightarrow \mathcal{R}^n$ which depends on an $n-$dimensional state-variable $x = \lbrace x_1, x_2, \ldots, x_n \rbrace$ and $m$ parameters which can be represented by a parameter vector $p = \lbrace p_1, p_2, \ldots, p_m \rbrace$.

Consider $y_i$ a component of the vector $y$. The value of each of $y_i$, in general, depends on the value of each parameter $p_j$ in the parameter vector $p$; and varying any parameter would result in a variation in $y_i$.The amount of variation caused in $y_i$ due to a unit variation in $p_j$ determines the sensitivity of $y_i$ to the parameter $p_j$. There are various well-known methods to quantify this sensitivity. In this paper, we have presented the results of variance-based sensitivity analysis introduced first by I. M. Sobol~\cite{Sobol2001}.

For a given component $y_i$, the variance-based sensitivity analysis gives indices $S_{i,j}$ corresponding to each parameter $p_j$. The index $S_{i,j}$ is known as the first order Sobol index corresponding to parameter $p_j$ and output component $y_i$. It correlates with the sensitivity of $y_i$ with respect to variations in $p_j$ and is defined as

\begin{equation}
    S_{i,j} = \frac{V_{i,j}}{V_i} = \frac{V \left( E \left(y_i | p_j \right) \right)}{V(y_i)},
    \label{eq:sobol}
\end{equation}

where $V_{i,j} = V \left( E \left(y_i | p_j \right) \right)$ is the variance in the expected value of $y_i$ when the parameter $p_j$ is fixed and $V_i = V(y_i)$ is the variance in the value of $y_i$ when no restrictions are imposed.

Note that the variation in $y$ may not only be caused by variation in single parameters, but also by variations in a combination of parameters. For example, in the results presented in the paper, if infection rate is varied and all other parameters are kept constant, noting the variation in daily number of hospitalizations gives us the first order Sobol index corresponding to infection rate. Similarly, if immunity duration varied and all other parameters are kept constant, we obtain the first order Sobol index corresponding to immunity duration. However, if infection rate and immunity duration are simultaneously varied, the variation in daily number of hospitalizations can be more than the sum of variations in the previous two cases. The measure of this additional variation in the output variable (daily number of hospitalizations) with respect to variation in more than one parameter gives rise to higher order Sobol indices. Investigations into these higher order Sobol indices can be a topic of future research.

It is clear from equation~\ref{eq:sobol} that the computation of Sobol indices requires variances in the expected values of the output variables. Since the output variables can only be defined implicitly when dealing with complex systems such as the one being studied in this paper, it becomes important to estimate this variance. To this end, various statistical techniques such as polynomial chaos expansion~\cite{Crestaux2009}, stochastic collocation~\cite{Eldred2009}, and quasi-Monte-Carlo method~\cite{Okten20211}. In this paper, we have used the stochastic collocation method for computing the variance of the output variables and hence the Sobol indices.

Stochastic collocation method essentially samples the parameter space using a number of discrete univariate points. The model is computed at these points in the parameter space. A multivariate interpolant is then constructed using the output of the model at these points. This interpolant is then used as an approximation of the output variable $y_i$ in equation~\ref{eq:sobol}. Further mathematical details of the stochastic collocation method can be seen in the literature~\cite{Tang2010}.

\subsection*{Software and computational facilities used}

\begin{figure}
    \centering
    \includegraphics[width=\textwidth]{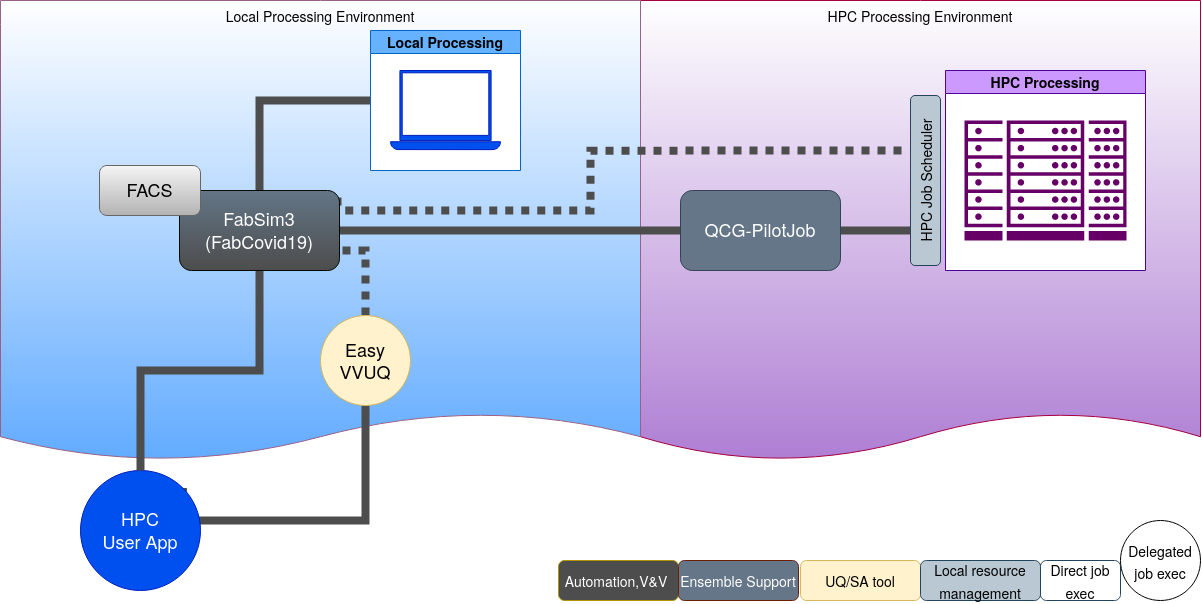}
    \caption{Overview of the FACS and FabCovid19 plugin in congestion with the SEAVEA toolkit. Lines represent the interaction between the components, where solid lines represent the default or most common interactions. Dashed lines represent the available alternatives.}
    \label{fig:tube}
\end{figure}

Given their computational requirements, the FACS simulations used in this paper were performed on ARCHER2 high-performance computers, which have a total of 5,860 nodes with an estimated peak performance of 28 Pflops per second.  Since ARCHER2 is used by a large number of researchers around the world, it is necessary to efficiently distribute the simulation between the HPC nodes. This was done using QCG-PilotJob~\cite{Bosak2021}. The sensitivity analysis and computation of the Sobol indices presented in the paper was performed using EasyVVUQ software~\cite{Richardson2020}. Simultaneous integration and use of the above-mentioned tools can become challenging. Therefore, we used FabSim3~\cite{Groen2023} as an interface, which gives access to these tools and HPC using one-line commands. Figure~\ref{fig:tube} shows the interaction between the various software components. We will now describe each of these components in some detail.

\subsubsection*{Simulation run time}

FACS is an open-source software written in Python, which can be run on any Linux machine. Since the code for FACS can run on multiple cores, the time required for each simulation run mostly depends on the number of agents in the simulation, the number of days to be simulated, and the number of cores being used by FACS. For the results presented in this paper, we have simulated regions with an approximate population of 200,000 people for a simulation period of 400 days. While running on a single core, a single run of such a simulation takes about an hour to run. In this paper, we presented sensitivity analysis results against a set of four parameters which define the disease vectors. As discussed earlier in the Results section, these parameters were selected out of seven possible scalar parameters based on preliminary sensitivity analysis. Taking into consideration the fact that each sensitivity analysis itself comprises an ensemble of 256 runs, a total of 2034 simulations were required to obtain these results. Out of these, results of 768 runs are presented in the paper. The total amount of time required to perform the required number of simulations would have been too large to be practically run on a personal computer. Therefore, we used ARCHER2, which is an HPC hosted at the University of Edinburgh. On ARCHER2, the simulations were run using 128 cores. Therefore, the amount of time required for each simulation run was reduced to under 60 seconds.

\subsubsection*{Computing the Sobol indices}

The computation of Sobol indices presented in this paper was handled by EasyVVUQ, a Python package to conduct verification, validation and uncertainty quantification for HPC simulations. One of the primary tools provided by this package is the computation of Sobol indices. Given a simulation which takes one or more input files, EasyVVUQ uses appropriately constructed templates to modify the numerical values present in the input files. The parameters to be varied, region of the parameter space to be scanned, the ensemble size and the sampling method to be used for the sensitivity analysis is described in a separate settings file. The ensemble of jobs is then created and submitted to the HPC. After the simulations have completed, the results are collected and Sobol indices are appropriately computed and stored in a database.

\subsubsection*{Interface with HPC}

Since HPCs are generally used only for running the simulations and not for analyzing or plotting the results, running a large ensemble of jobs on the HPC would normally be a cumbersome process which would involve organizing the input and output data, creating the job submission scripts, job submission, transferring the output data back to the local machine and further post-processing. Performing all these steps manually would not only be time-consuming but also prone to errors. Hence, we use FabSim3, a software written in Python based on the Fabric2 framework which provides an interactive user interface to the HPCs.

FabSim3 essentially automates the above-mentioned workflow by preparing shell scripts and executing a set of commands automatically. For instance, as mentioned earlier, the computation of Sobol indices involves running a large ensemble of jobs which are prepared using the EasyVVUQ package. FabSim3 allows us to prepare the ensembles and submit the jobs using a single command issued on the local machine. A second command can then transfer the output of the simulations back to the local machine, compute the Sobol indices and prepare the plots shown in Figure~\ref{fig:sobol}. Hence, the Sobol indices can be computed using two commands in total. FabSim3 has also been used to prepare all other plots in the paper too. 

The computation and visualization tools specific to FACS are handled through a FabSim3 plugin called FabCovid19. These plugins use the general-purpose API's offered by FabSim3 for a specific software. Using such plugins, FabSim3 has also been applied to various other computationally intensive simulations in other fields of research~\cite{Daub2021,Wright2020}.

\bibliography{sample}



\section*{Acknowledgements}
This work has been supported by the SEAVEA ExCALIBUR project, which has received funding from EPSRC under grant agreement EP/W007711/1, as well as by the STAMINA project, which has received funding from the European Union Horizon 2020 research and innovation programme under grant agreement no 883441. The simulations were performed using the ARCHER2 UK National Supercomputing Service (Project code: e723).

\section*{Author contributions statement}

A.S. conducted the simulations and wrote the manuscript. M.G. prepared the parts of the input parameters. D.S. revised the mathematical aspects of the analysis. D.G. and A.A. conceived the project and supervised the work. All authors reviewed the manuscript.


\section*{Additional information}

The authors declare no competing interests.






\end{document}